\documentclass[twocolumn,english,aps,prl,showpacs,superscriptaddress]{revtex4}
\usepackage[T1]{fontenc}
\usepackage[latin9]{inputenc}
\usepackage{amstext}
\usepackage{graphicx}
\usepackage{color}
\usepackage{amsmath}
\usepackage{amssymb}
\makeatletter

\@ifundefined{textcolor}{}
{%
 \definecolor{BLACK}{gray}{0}
 \definecolor{WHITE}{gray}{1}
 \definecolor{RED}{rgb}{1,0,0}
 \definecolor{GREEN}{rgb}{0,1,0}
 \definecolor{BLUE}{rgb}{0,0,1}
 \definecolor{CYAN}{cmyk}{1,0,0,0}
 \definecolor{MAGENTA}{cmyk}{0,1,0,0}
 \definecolor{YELLOW}{cmyk}{0,0,1,0}
 }


\newcommand{\1}{{\rm 1\hspace{-0.9mm}l}}
\newcommand{\Id}{\1}

\makeatother

\begin{document}

\title{{ 
Constructive entanglement test from triangle inequality}}

\author{\L{}ukasz Rudnicki}

\email{rudnicki@cft.edu.pl}

\affiliation{Center for Theoretical Physics, Polish Academy of Sciences, Aleja
Lotnik{\'o}w 32/46, PL-02-668 Warsaw, Poland}
\affiliation{Freiburg Institute for Advanced Studies, Albert-Ludwigs University
of Freiburg, Albertstrasse 19, 79104 Freiburg, Germany }

\author{Zbigniew Pucha\l{}a}

\affiliation{Institute of Theoretical and Applied Informatics, Polish Academy
of Sciences, Ba\l{}tycka 5, PL-44-100 Gliwice, Poland}

\affiliation{Smoluchowski Institute of Physics, Jagiellonian University, ul. Reymonta
4, PL-30-059 Krak{\'o}w, Poland}

\author{Pawe\l{} Horodecki}

\affiliation{Faculty of Applied Physics and Mathematics, Technical University
of Gda\'{n}sk, PL-80-952 Gda\'{n}sk, Poland}

\affiliation{National Quantum Information Centre of Gda\'{n}sk, PL-81-824 Sopot,
Poland}

\author{Karol \.{Z}yczkowski}

\affiliation{Center for Theoretical Physics, Polish Academy of Sciences, Aleja
Lotnik{\'o}w 32/46, PL-02-668 Warsaw, Poland}

\affiliation{Smoluchowski Institute of Physics, Jagiellonian University, ul. Reymonta
4, PL-30-059 Krak{\'o}w, Poland}


\begin{abstract}
We derive a simple lower bound on the geometric
measure of entanglement for mixed  quantum states in the case of a general
multipartite system. The main ingredient of the presented
derivation is the triangle inequality applied to the root infidelity distance in the space of density matrices.
The obtained bound leads to entanglement criteria
with a straightforward interpretation. Proposed criteria provide an experimentally accessible, powerful entanglement test.
\end{abstract}
\maketitle

Quantum entanglement characterizes non--classical correlations
in a quantum system consisting of several subsystems \cite{Schrodinger,Einstein,Bengtsson,Horodeccy,PR_Guhne}.
In the case of a pure quantum state, any correlations
between subsystems, that can be detected in coincidence experiments,
confirm entanglement. However, in any realistic experiment one has to cope
with mixed quantum states, for which the problem becomes more involved,
as quantum and classical correlations may exist. To detect reliably 
quantum entanglement for a mixed quantum state one needs to rule out
the more common case of classical correlations.

While efficient detection of quantum entanglement
is not an easy task in quantum information theory, 
it is  more difficult to characterize this phenomenon 
quantitatively basing on results of partial measurements
 that are not sufficient for full state reconstruction.
Known schemes of such experimental procedure require 
interactions between many copies of the state investigated \cite{PH}.
With an interaction between two copies of the state
one can estimate a lower bound on an entanglement 
measure \cite{mintert} 
in terms of a  two--copy entanglement witness 
that reproduces the difference between global and local entropy \cite{PH1}. 
The things become more complicated 
in the case of the restriction to single-copy measurements. 
Despite some recent progress (see \cite{Eberly})
still there is no general satisfactory answer to the question 
how well one can estimate given entanglement measure 
on the basis of noncomplete (ie. non--tomographic) experimental data.



In this work we build on a pragmatic approach advocated in
\cite{AL09,PR_Guhne}, in which one attempts to construct 
entanglement measures accessible in an experiment.
We derive a lower bound for the \textit{geometric measure of entanglement} (GME)
\cite{Brody,geometric}  capable to describe entanglement of an arbitrary mixed quantum state. We shall emphasize that we are
unaware of any results concerning lower bounds for GME (an upper bound given in terms of the generalized robustness of entanglement can be found in \cite{Cavalcanti}). 
We demonstrate that our quantity can be used to compare the amount of entanglement between different 
 mixed states of a $d\times d$ system
and provides a separability test which is experimentally accessible. 


Consider an arbitrary $K$--partite quantum system described in the Hilbert space
$\mathcal{H}=\bigotimes_{I=1}^K\mathcal{H}^{I}$ with no assumption about
the dimensionality of the particular subspace $\mathcal{H}^{I}$ representing
the $I$-th subsystem. 
We denote by 
$\mathcal{S}_{m}$, $m=1,\ldots,K$
the set of $m$--separable pure states $\left|\phi\right\rangle _{\textrm{sep}}^m=\otimes_{I=1}^m\left|\phi_{I}\right\rangle $. We have the following chain $S_K\subset S_{K-1}\subset\cdots\subset S_1=\mathcal{H}$.

In our considerations we shall use the root infidelity distance between
two mixed states $\rho_{1}$ and $\rho_{2}$ \cite{rootinfidel}:
\begin{equation}
C_{F}\left(\rho_{1},\rho_{2}\right)=\sqrt{1-F\left(\rho_{1},\rho_{2}\right)},\label{infidelity}
\end{equation}
defined with the help of the fidelity $F\left(\rho_{1},\rho_{2}\right)$.
To derive our result we use the fidelity involving
at least one pure state, thus we need only the restricted, simpler
formula for the fidelity 
\begin{equation}
F\left(\rho,\left|\Psi\right\rangle \left\langle \Psi\right|\right)\equiv F\left(\rho,\left|\Psi\right\rangle\right)=\left\langle \Psi\right|\rho\left|\Psi\right\rangle .\label{fidelity}
\end{equation}

Finally we need to introduce the hierarchy of GME \cite{Brody,geometric,Cavalcanti}, 
which in the case of pure states
is defined as:
\begin{eqnarray}
E_m\left(\left|\Psi\right\rangle \right) & = & 1-\max_{\left|\phi\right\rangle \in\mathcal{S}_{m}}\left|\!\left.\left\langle \phi\right|\Psi\right\rangle \right|^{2}\label{geom meas}\\
 & \equiv & \min_{\left|\phi\right\rangle
 \in\mathcal{S}_{m}}C_{F}^{2}\left(\left|\phi\right\rangle 
  ,\left|\Psi\right\rangle \right),
\quad m=2,\dots, K .\nonumber 
\end{eqnarray}
The second, equivalent definition follows directly from Eqs. (\ref{infidelity},
\ref{fidelity}). The operational interpretation of the measure $E_m\left(\left|\Psi\right\rangle \right)$
is straightforward. If the state $\left|\Psi\right\rangle $ is $m$--separable
it belongs to the set $\mathcal{S}_{m}$, thus the minimal infidelity
distance is $0$, since one can always chose $\left|\phi\right\rangle \in\mathcal{S}_{m}$
to be equal $\left|\Psi\right\rangle $. 

The geometric measure of entanglement for mixed states
is defined \cite{geometric} with the help of the convex roof construction:
\begin{equation}
E_m\left(\rho\right)=\min_{\mathcal{E}}\sum_{i}p_{i}\min_{\left|\phi\right\rangle \in\mathcal{S}_{m}}C_{F}^{2}\left(\left|\phi\right\rangle ,\left|\Psi_{i}\right\rangle \right),\label{mixed1}
\end{equation} where the ensemble $\mathcal{E}=\left\{ p_{i},\left|\Psi_{i}\right\rangle \right\} $
represents the mixed state $\rho$, i.e. $\rho=\sum_{i}p_{i}\left|\Psi_{i}\right\rangle \left\langle \Psi_{i}\right|$.
Surprisingly, it was shown \cite{strelsov} that $E_m\left(\rho\right)$
is simultaneously a distance measure $E_m\left(\rho\right)=\min_{\sigma}C_{F}^{2}\left(\sigma,\rho\right)$ with $\sigma$ being a $m$--separable mixed state, for $m=2,\dots,K$.

\paragraph{The lower bound on $E_m\left(\rho\right)$.---} 
Any density matrix representing a multipartite system 
can be characterized by its {\sl product numerical radii} $L_m(\rho)$,
often used in the theory of quantum information  \cite{Gaw+10}.
These quantities can be defined as 
the maximal expectation value of $\rho$ among normalized pure product states,
\begin{equation}
L_m\left(\rho\right)=\max_{\left|\phi\right\rangle 
\in\mathcal{S}_{m}}\left\langle \phi\right|\rho\left|\phi\right\rangle
, \quad m=2,\dots K
 .\label{L}
\end{equation}
Note that $E_m\left(\left|\Psi\right\rangle \right)\equiv1-L_m\left(\left|\Psi\right\rangle \right)$. 

The main result of this paper is the following lower bound for
the square root of the geometric measures of entanglement
\begin{equation}
\sqrt{E_m\left(\rho\right)}\geq \mathcal{R}_m\left(\rho\right)=\sqrt{1-L_m\left(\rho\right)}-\sqrt{1-\textrm{Tr}\rho^{2}}.\label{Lower bound}
\end{equation}

We start the derivation of (\ref{Lower bound}) with an arbitrary
expansion $\rho=\sum_{i}p_{i}\left|\Psi_{i}\right\rangle \left\langle \Psi_{i}\right|$
of the mixed state $\rho$. For some fixed index $i$ we chose a pure
state $\left|\Psi_{i}\right\rangle $, and another pure state $\left|\phi\right\rangle $
to be specified. Since the root infidelity (\ref{infidelity}) is
a legitimate metric we can write down the triangle inequality for
$C_{F}\left(\left|\phi\right\rangle ,\rho\right)$ with $\left|\Psi_{i}\right\rangle \left\langle \Psi_{i}\right|$
as a third state:
\begin{equation}
C_{F}\left(\left|\phi\right\rangle ,\rho\right)\leq C_{F}\left(\left|\phi\right\rangle ,\left|\Psi_{i}\right\rangle \right)+C_{F}\left(\left|\Psi_{i}\right\rangle ,\rho\right).
\end{equation}
If we next take the minimum with respect to $\left|\phi\right\rangle \in\mathcal{S}_{m}$
and use the definitions (\ref{infidelity}, \ref{geom meas}, \ref{L})
we obtain
\begin{equation}
\sqrt{1-L_m\left(\rho\right)}\leq\sqrt{E_m\left(\left|\Psi_{i}\right\rangle \right)}+C_{F}\left(\left|\Psi_{i}\right\rangle ,\rho\right).\label{intermediate}
\end{equation}
In the next step we shall multiply the resulting inequality by $p_{i}$
and sum over $i$. The term $\sqrt{1-L_m\left(\rho\right)}$ is independent
of $i$, while for the two terms on the right hand side we shall apply
the following estimates originating from the concavity of the $\sqrt{\cdot}$
function:
\begin{equation}
\sum_{i}p_{i}\sqrt{E_m\left(\left|\Psi_{i}\right\rangle \right)}\leq\sqrt{\sum_{i}p_{i}E_m\left(\left|\Psi_{i}\right\rangle \right)},\label{estim1}
\end{equation}
\begin{equation}
\sum_{i}p_{i}C_{F}\left(\left|\Psi_{i}\right\rangle ,\rho\right)\leq\sqrt{1-\sum_{i}p_{i}\left\langle \Psi_{i}\right|\rho\left|\Psi_{i}\right\rangle }.\label{estim2}
\end{equation}
In the final step we shall recognize that the sum over $i$ on the
right hand side of (\ref{estim2}) is equal to $\textrm{Tr}\rho^{2}$,
so that is independent of the given ensemble $\mathcal{E}=\left\{ p_{i},\left|\Psi_{i}\right\rangle \right\} $.
This implies that we can immediately minimize with respect to $\mathcal{E}=\left\{ p_{i},\left|\Psi_{i}\right\rangle \right\} $
 producing the quantity $\sqrt{E_m\left(\rho\right)}$ on the right hand side of (\ref{estim1}).
Applying the above estimates to Eq. (\ref{intermediate}) we
obtain the desired lower bound (\ref{Lower bound}) after a one--step rearrangement.
 From (\ref{Lower bound}) one can obviously find the lower bound for $E_m\left(\rho\right)$,
which reads $\left(\max\left[\mathcal{R}_m\left(\rho\right);0\right]\right)^{2}$.
It is important to take the maximum first, in order to avoid cases
when negative values of $\mathcal{R}_m\left(\rho\right)$ can give a positive,
unphysical contribution $\mathcal{R}_m^{2}\left(\rho\right)$ to the lower bound of $E_m\left(\rho\right)$.

We shall further observe that the quantity $1-L_m\left(\rho\right)$ provides a natural 
(but typically rough)
upper bound for $E_m\left(\rho\right)$,
To prove this statement it is sufficient to restrict the minimization in $E_m\left(\rho\right)=\min_{\sigma}C_{F}^{2}\left(\sigma,\rho\right)$ to pure states $\sigma=\left|\phi\right\rangle \left\langle\phi\right|$. This upper bound, as well as our lower bound are in the case of pure states equal to  $E_m\left(\rho\right)$. In the case of \textit{pseudo--pure} states characterized by $\textrm{Tr}\rho^{2}\lesssim1$ we thus get a sharp estimate of the value of the entanglement measure in question.

\begin{figure}
\includegraphics[scale=0.45]{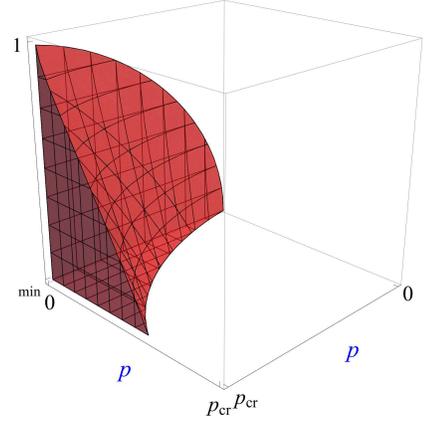}\caption{(color online).
Parameter space for the generalized Werner states of a $3 \times 3$ system.
Red volume corresponds to the states $\rho_{p',\boldsymbol{\lambda}}$ 
the entanglement of which is shown by the bound (\ref{Lower bound}) to be larger than this of reference state $\rho_{p,\boldsymbol{\bar{\lambda}}}$.
Here $d=3$ so that $p_{\textrm{cr}}=3/4$ and $\Lambda_{\textrm{min}}=1/3$.
}
\label{Fig:Detection}
\end{figure}

Let us now focus on the family of generalized $d\times d$ Werner states ($0\leq p\leq1$):
\begin{equation}
\rho_{p,\boldsymbol{\lambda}}=\left(1-p\right)\left|\Psi_{\boldsymbol{\lambda}}\right\rangle \left\langle \Psi_{\boldsymbol{\lambda}}\right|+p\frac{I_d\otimes I_d}{d^2},\label{WST}
\end{equation}
where $\left|\Psi_{\boldsymbol{\lambda}}\right\rangle =\sum_{i=1}^d\sqrt{\lambda_i}\left|ii\right\rangle$. One can straightforwardly calculate (as $\rho_{p,\boldsymbol{\lambda}}$ is bipartite we can skip the index $m$):
\begin{equation}
\textrm{Tr}\rho_{p,\boldsymbol{\lambda}}^{2}=1+\frac{\left(p^2\!-\!2p\right)\left(d^2\!-\!1\right)}{d^2},\;\; L\left(\rho_{p,\boldsymbol{\lambda}}\right)=\frac{p}{d^2}+\left(1-p\right)\Lambda, \label{LWerner}
\end{equation}
where $\Lambda=\max_i\lambda_i$ and $1/d\leq \Lambda\leq 1$. 
The above family possesses a distinguished member given by  $\boldsymbol{\bar{\lambda}}=\left(1/d,\ldots,1/d\right)$, which for $p=0$ represents the maximally entangled state. 
Comparing $E\left(\rho_{p,\boldsymbol{\bar{\lambda}}}\right)$ with the lower bound based on $\mathcal{R}\left(\rho_{p',\boldsymbol{\lambda}}\right)$ for $p'<p$ and 
vector $\boldsymbol{\lambda}$ majorizing  \cite{MO79}
 $\boldsymbol{\bar{\lambda}}$
we can look for the states $\rho_{p',\boldsymbol{\lambda}}$ which are more entangled than $\rho_{p,\boldsymbol{\bar{\lambda}}}$ (see Fig. \ref{Fig:Detection} and \cite{Supplement}). 


\paragraph{Entanglement criteria.---}
Eq. (\ref{Lower bound}) provides the entanglement criteria
\begin{equation}
\left(L_m\left(\rho\right)<\textrm{Tr}\rho^{2}\right)\;\Rightarrow\;\left(\rho\textrm{ is not m--separable}\right),\label{Criteria}
\end{equation}
which for $m=K$ have been recently recognized in
\cite{GSarbicki,Badziag}. In \cite{GSarbicki} they appear in the form of a nonlinear entanglement witness $L_K\left(\rho\right)\Id-\rho$,
while in \cite{Badziag} a more general object (see Eq. (12) from
\cite{Badziag}) that contains $L_K\left(\rho\right)$ as a special case
has been introduced. 
The above situation is similar to the case of \textit{purity--entropy} entanglement criteria \cite{pur/ent}
given in terms of the R\'enyi entropy $H_{\alpha}$, which for $\alpha=2$ was shown \cite{mintert} to establish the lower
bound for the concurrence \cite{concurence}. Let us emphasize in passing that the criteria (\ref{Criteria})
are strong enough to detect bound entanglement \cite{Supplement} of a concrete family \cite{Bound-activation}. 

Let us now study the entanglement criteria (\ref{Criteria}) in the case of a bipartite $M\times N$ system. A general mixed state of such system can be written as
\begin{eqnarray}
\rho & = & \frac{1}{MN}\left[I_{M}\otimes I_{N}+ k_M \sum_{i=1}^{M^{2}-1}q_{i}\sigma_{i}\otimes I_{N}\right.\\
 & + & \left. k_N \sum_{j=1}^{N^{2}-1}p_{j}I_{M}\otimes\tilde{\sigma}_{j}+k_M k_N \sum_{i=1}^{M^{2}-1}\sum_{j=1}^{N^{2}-1}B_{ij}\sigma_{i}\otimes\tilde{\sigma}_{j}\right],\nonumber 
\end{eqnarray}
where $\sigma_{i}$, $i=1,\ldots,M^2-1$ and $\tilde{\sigma}_j$,  $j=1,\ldots,N^2-1$ are traceless, hermitian generators of $SU\left(M\right)$ and $SU\left(N\right)$ groups respectively,
normalized such that $\textrm{Tr}\sigma_{i}\sigma_{i'}=2\delta_{i'i}$ and $\textrm{Tr}\tilde{\sigma}_{j}\tilde{\sigma}_{j'}=2\delta_{j'j}$. For further convenience we set $k_M=\sqrt{M M_-/2}$ where $M_-=M-1$ and similarly for $k_N$.
In the above representation the state is described by two Bloch vectors $\boldsymbol{p}$,  $\boldsymbol{q}$ of the partially reduced states and the $\left(M^2-1\right)\times\left(N^2-1\right)$ correlation tensor $\mathbf{B}$ 
\cite{Bengtsson}.
Let us recall that the Bloch vector $\boldsymbol{q}$ belongs to the space $\mathcal{B}\left(M\right)$ defined by the constraints   $\boldsymbol{q}\cdot\boldsymbol{q}=1$ and $2\left(M-2\right)\boldsymbol{q}=k_{M}\textrm{Tr}\left(\left(\boldsymbol{q}\cdot\boldsymbol{\sigma}\right)^{2}\boldsymbol{\sigma}\right)$ \cite{Bloch1, Bloch2}, and an equivalent definition holds for  $\boldsymbol{p}\in\mathcal{B}\left(N\right)$.

The pure separable state $\left|\phi\right\rangle \left\langle \phi\right|$ present in (\ref{L}) can be completely characterized by a couple of Bloch vectors $\boldsymbol{v}\in\mathcal{B}\left(M\right)$ and $\boldsymbol{w}\in\mathcal{B}\left(N\right)$. In that representation the product numerical radius $L\left(\rho\right)$ reads 
\begin{equation}
\frac{1+\max_{\boldsymbol{v},\boldsymbol{w}}\left(M_{-}\boldsymbol{v}\!\cdot\boldsymbol{q}+N_{-}\boldsymbol{w}\!\cdot\boldsymbol{p}+M_{-}N_{-}\boldsymbol{v}\!\cdot\!\mathbf{B}\boldsymbol{w}\right)}{MN}. \label{numer}
\end{equation}
The above maximization over $\left(\boldsymbol{v},\boldsymbol{w}\right)\in\mathcal{B}\left(M\right)\times\mathcal{B}\left(N\right)$ can be efficiently performed numerically even for larger systems, eg. $M,N=10$. To get however a deeper insight we provide in the Supplemental Material \cite{Supplement} the following upper bound for (\ref{numer}):
\begin{equation}
L\left(\rho\right)\leq\frac{1+N_{-}\left\Vert \boldsymbol{p}\right\Vert +M_{-}\left\Vert \boldsymbol{q}\right\Vert +M_{-}N_{-}\sqrt{\xi_{1}\left(\mathbf{C}\right)}}{MN},\label{boundL}
\end{equation}
where $\xi_{1}\left(\mathbf{C}\right)$ denotes the largest eigenvalue of $\mathbf{C}=\mathbf{B}^T\mathbf{B}$ or $\mathbf{C}=\mathbf{B}\mathbf{B}^T$ (both quadratic matrices  $\mathbf{B}^T\mathbf{B}$ and $\mathbf{B}\mathbf{B}^T$ possess the same trace and nonzero eigenvalues).  The purity of $\rho$ can also be easily computed:
\begin{equation}
\textrm{Tr}\rho^{2}=\frac{1+N_{-}\left\Vert \boldsymbol{p}\right\Vert ^{2}+M_{-}\left\Vert \boldsymbol{q}\right\Vert ^{2}+M_{-}N_{-}\textrm{Tr}\mathbf{C}}{MN}.
\end{equation}
In fact, if any upper bound on  $L_m\left(\rho\right)$ is smaller than $\textrm{Tr}\rho^{2}$ then the condition (\ref{Criteria}) is satisfied. In the above case our entanglement test can thus be rewritten as: If
\begin{equation}
M_-\left\Vert \boldsymbol{q}\right\Vert \left(1-\left\Vert \boldsymbol{q}\right\Vert \right)+N_-\left\Vert \boldsymbol{p}\right\Vert \left(1-\left\Vert \boldsymbol{p}\right\Vert \right)+M_-N_-f\left(\mathbf{C}\right)<0,\label{CritNew}
\end{equation}
 where
$f\left(\mathbf{C}\right)=\sqrt{\xi_{1}\left(\mathbf{C}\right)}-\textrm{Tr}\mathbf{C}$,
then the mixed state $\rho$ is entangled.  Note that the above criterion 
is invariant under local unitary operations 
$U_{A} \otimes U_{B}$ what follows from the fact that a unitary rotation $UXU^{\dagger}$ 
of any matrix $X$  is an isometry in the Hilbert-Schmidt space. 
This means that for a given state $\rho_{AB}$ entanglement of its all 
$U_{A} \otimes U_{B}$ transformations is detected with the same efficiency.

In order to investigate the performance of the new criteria (\ref{CritNew}) we shall use the state $\rho_{p,\lambda}\equiv\rho_{p,\boldsymbol{\lambda}}$ given by Eq. (\ref{WST}) with $d=2$, so that $\boldsymbol{\lambda}=\left(\lambda,1-\lambda\right)$.
This state is described by $\boldsymbol{p}=\boldsymbol{q}=\left(0,0,z\right)$, with $z=\left(1-p\right)\left(2\lambda-1\right)$, and $\mathbf{B}=\left(1-p\right)\textrm{diag}\left(\eta,\eta,1\right)$ with $\eta=2\sqrt{\lambda\left(1-\lambda\right)}$. By a direct substitution and comparison with (\ref{LWerner}) one can check that the bound (\ref{boundL}) becomes tight.

From the PPT criteria we know that $\rho_{p,\lambda}$ is separable when $p\geq1-\left(1+2\eta\right)^{-1}$. In the maximally entangled case $\lambda=1/2$ the threshold for separability is thus $p_{\textrm{sep}}=2/3$.
According to the criteria (\ref{CritNew}) the state $\rho_{p,\lambda}$ is
entangled for $p\leq4\left(1-\max\left[\lambda;1-\lambda\right]\right)/3$.
Note that for $\lambda=1/2$ we obtain the separability threshold
$p=2/3$, so that a full range of entangled Werner states
is detected. This conclusion remains valid for an arbitrary dimension $d$, where \cite{deVic}  $\boldsymbol{p}=0=\boldsymbol{q}$ and $\mathbf{C}$ is the identity matrix multiplied by $\left(1-p\right)^2/\left(d-1\right)^2$.
\begin{figure}
\includegraphics[scale=0.34]{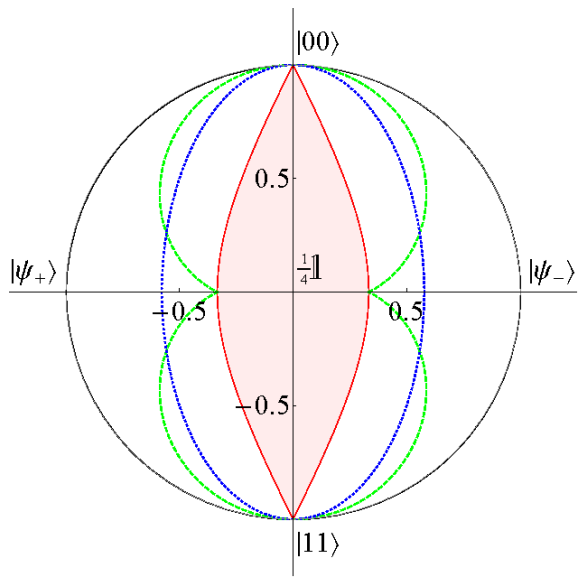}\caption{(color online). The plane of
a family of two--qubit states $\rho_{p,\lambda}$ (points inside gray
circle) in the polar coordinate $r=1-p$, $\theta=2\arccos\sqrt{\lambda}$
. States in the shaded region bounded by solid red curve are separable
(PPT). Entangled states outside dashed green curve and outside dotted
blue curve are detected by the criteria (\ref{CritNew}) and the purity
test respectively.\label{Fig:Werner}}
\end{figure}

In Fig. \ref{Fig:Werner} we compare the criteria (\ref{CritNew})
(dashed green curve) and the purity--entropy test \cite{pur/ent} (dotted
blue curve) given by the condition $\textrm{Tr}\rho_{A/B}^{2}\geq\textrm{Tr}\rho^{2}$
satisfied when $\rho$ is separable. Here $\rho_{A/B}$ denote density
operators of single subsystem $A$ and $B$ respectively. In the neighborhood of the maximally entangled state ($\lambda=1/2$)
the criteria (\ref{Criteria}) outperform the purity test.

The above criteria has somehow built in the purity requirement but, as we 
have seen above, its power does not necessarily depend on how pure the state is.
It might however be sensitive to the degree of  purity of the element in the mixture 
that is responsible for entanglement. To study this possibility we check the family of $U \otimes U$ invariant $ d \otimes d$ original Werner states \cite{Werner89}. This family is defined by $\rho(\alpha,d)=\left(I_{d} \otimes I_{d} + \alpha V\right)/\left(d^{2} + \alpha d\right)$ with the swap operator $V$, a real parameter $\alpha \in [-1,1]$ and sharp entanglement condition $\alpha\in [-1,-1/d)$ following from PPT test. 
The states are known to have $\boldsymbol{p}=0=\boldsymbol{q}$, what can be seen immediately by considering their 
partial transpose. With the help of formula  $\textrm{Tr}(A_1 \otimes A_2 V)=\textrm{Tr}(A_1A_2)$ \cite{Werner89}  the matrix $\mathbf{B}=\alpha d I_d/ (d^{2} + \alpha d)(d-1)$ can be 
easily found. The condition (\ref{CritNew}) 
reports entanglement for $\alpha \in [-1,-\frac{d}{d+2})$ converging to the single-point extreme $\alpha=-1$
rather than the
entanglement-corresponding interval $[-1,-1/d)$ with $d \rightarrow \infty $. Using Eq. (\ref{numer}) we see however that
\begin{equation}
(d-1)^2\max_{\boldsymbol{v},\boldsymbol{w}}\boldsymbol{v}\cdot\mathbf{B}\boldsymbol{w}=\frac{\alpha\left(d-1\right)}{d+\alpha}\max_{\boldsymbol{v},\boldsymbol{w}}\boldsymbol{v}\cdot\boldsymbol{w}=-\frac{\alpha}{d+\alpha},
\end{equation}
provided that $\alpha<0$. The last equality appears since if $\boldsymbol{v},\boldsymbol{w}\in \mathcal{B}\left(d\right)$ then $-1/(d-1)\leq\boldsymbol{v}\cdot\boldsymbol{w}\leq1$ \cite{Bloch1}.
With the above result we recover the entanglement condition $\alpha<-1/d$.

\paragraph{Experimental advantages.---}
The entanglement test (\ref{CritNew}) can be successfully used provided that $\boldsymbol{p}$,
$\boldsymbol{q}$ and $\mathbf{C}$ are known. In fact, in order to
determine $\xi_{1}\left(\mathbf{C}\right)$ all matrix elements
of the positive, symmetric matrix $\mathbf{C}$ must be found, what
in principle requires quantum tomography. In the two--qubit case,
our criteria while faithful on the family of Werner states, will always
be less practical than the PPT condition. Our aim is thus to reduce
the number of necessary parameters. To this end we shall upper bound
the function $f\left(\mathbf{C}\right)$, so that the upper bound
$g\left(\mathbf{C}\right)\geq f\left(\mathbf{C}\right)$ depends on
less number of matrix entries. If inequality (\ref{CritNew}) is satisfied with
$f\left(\mathbf{C}\right)$ substituted by $g\left(\mathbf{C}\right)$
then the state in question is obviously entangled. In order to achieve
this goal we shall distinguish the matrix elements of $\mathbf{C}$ to be measured and maximize $f\left(\mathbf{C}\right)$ with respect
to the remaining parameters. Performing the maximization we shall
preserve the positivity of $\mathbf{C}$. 

Let us explain the above approach using an example of two qubits,
so that $\mathbf{C}$ is a real, symmetric, $3\times3$ matrix given
by six parameters: $C_{11}$, $C_{12}$, $C_{22}$, $C_{13}$, $C_{23}$,
$C_{33}$. Assume that we would like to measure $C_{11}$, $C_{12}$,
$C_{22}$ and optimize $f\left(\mathbf{C}\right)$ with respect to
$C_{13}$, $C_{23}$, $C_{33}$. We obtain: 
\begin{equation}
f\left(\mathbf{C}\right)\leq g\left(C_{11},C_{12},C_{22}\right)\equiv\max_{C_{13},C_{23},C_{33}}f\left(\mathbf{C}\right).
\end{equation}

For the two--qubit Werner state we have: $C_{11}=\left(1-p\right)^{2}=C_{22}$,
$C_{12}=0$, so that the condition for $\mathbf{C}$ to be positive
reads:
\begin{equation}
C_{13}^{2}+C_{23}^{2}\leq C_{33}\left(1-p\right)^{2}.
\end{equation}
After analytical optimization \cite{Supplement} we find $g\left(C_{11},C_{12},C_{22}\right)\equiv g\left(p\right)$
of the form:
\begin{equation}
g\left(p\right)=\begin{cases}
\left(1-p\right)\left(2p-1\right) & \qquad\textrm{for }0\leq p\leq\frac{1}{2}\\
\frac{1}{4}-\left(1-p\right)^{2} & \qquad\textrm{for }\frac{1}{2}\leq p\leq1
\end{cases}.
\end{equation}
According to the test (\ref{CritNew}) the Werner state is detected to be entangled
if $g\left(p\right)<0$, so that for $p<1/2$. This is up to now the
best known threshold value for entanglement verification of the family of the Werner
states, obtained without resorting to quantum tomography. Let
us remind that the threshold value given by the purity test is $p=1-1/\sqrt{3}\approx0.4226$.

In fact, $10$ parameters suffice (see the Supplemental Material \cite{Supplement} for
explicit relations between the desired and measured parameters) to
determine $\left\Vert \boldsymbol{p}\right\Vert $, $\left\Vert \boldsymbol{q}\right\Vert $
and $C_{11}$, $C_{12}$, $C_{22}$. It is once again a huge experimental
advantage, as in order to measure the global purity $\textrm{Tr}\rho^{2}$
of a two--qubit state one needs $12$ parameters. This improvement
could be obtained because of the interplay between the purity $\textrm{Tr}\rho^{2}$
and the product numerical radius $L\left(\rho\right)$.

At the end let us analyze the $K$--qubit state
\begin{equation}
\varrho(p)=\left(1-p\right)\left|\Phi_{K}\right\rangle \left\langle \Phi_{K}\right|+\frac{p}{2^K} I^{\otimes K},
\end{equation}
where $\left|\Phi_{K}\right\rangle =\left(\left|0\right\rangle^{\otimes K} +\left|1\right\rangle^{\otimes K}\right)/\sqrt{2} $. The high symmetry of the above state provides that  $L_m\left(\varrho\right)=\left(1-p\right)/2+p/2^K$ for all $m\in\left\{2,\ldots,K\right\}$, what implies that all bounds $R_m\left(\varrho\right)$ capture the \textit{genuine multipartite entanglement} of $\varrho$ associated with $m=2$. In fact, $R_m\left(\varrho\right)=0$  leads to the biseparability threshold $p_{\textrm{gme}}=1/2\left(1-2^{-K}\right)$, which according to \cite{Seev} is optimal. 
\begin{acknowledgments} It is a great pleasure to thank Florian Mintert for his fruitful comments. This research was supported by the grant number IP2011 046871 ({\L}.R.) of the Polish Ministry of Science and Higher Education, and the grants number: DEC--2012/04/S/ST6/00400 (Z.P.) and \nopagebreak  2011/02/A/ST2/00305  \nopagebreak  (K.\.Z.) financed by Polish National Science Centre. A partial support from  EC\nopagebreak through the project Q--ESSENCE (P.H.)  is gratefully acknowledged.
\end{acknowledgments}

\section{Entanglement ordering}
\begin{figure}
\includegraphics[scale=0.5]{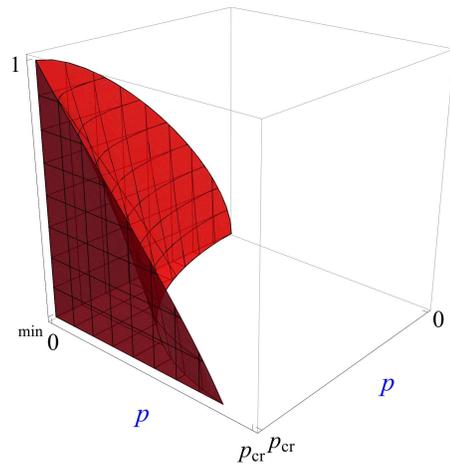}\caption{Parameter space for the generalized Werner states of a $10 \times 10$ system.
Red volume corresponds to the states $\rho_{p',\boldsymbol{\lambda}}$ 
the entanglement of which is shown by our bound to be larger than this of reference state $\rho_{p,\boldsymbol{\bar{\lambda}}}$. Here $d=10$ so that $p_{\textrm{cr}}=10/11$ and $\Lambda_{\textrm{min}}=1/10$.
\label{Fig:Detection1}}
\end{figure}
In \cite{geometric} it was shown that for the state $\rho_{p,\boldsymbol{\bar{\lambda}}}$ one can find the exact formula for 
the geometric measure of entanglement (GME):
\begin{equation}
E\left(\rho_{p,\boldsymbol{\bar{\lambda}}}\right)=1-\frac{1}{d}\left(\mathcal{\sqrt{F}}+\sqrt{\left(d-1\right)\left(1-\mathcal{F}\right)}\right)^{2},
\end{equation}
where
\begin{equation}
\mathcal{F}=1-\frac{p\left(d^{2}-1\right)}{d^{2}}.
\end{equation}
Using Eq. (12) of our paper we can find that
\begin{equation}
\mathcal{R}\left(\rho_{p',\boldsymbol{\lambda}}\right)=
\sqrt{1-\frac{p'}{d^2}-\left(1-p'\right)\Lambda}-\frac{\sqrt{\left(2p'\!-\!\left(p'\right)^2\right)\left(d^2\!-\!1\right)}}{d} .
\end{equation} 
The relation  $\mathcal{R}\left(\rho_{p',\boldsymbol{\lambda}}\right)>E\left(\rho_{p,\boldsymbol{\bar{\lambda}}}\right)$ is a sufficient condition for $\rho_{p',\boldsymbol{\lambda}}$ to be more entangled than $\rho_{p,\boldsymbol{\bar{\lambda}}}$, which is considered to be our reference state.

In the main paper we present the first nontrivial case of $d=3$. 
The volume of states found by our lower bound for $d=10$
is shown in Fig. \ref{Fig:Detection1}. 

\section{Bound entanglement} Consider now an example of the two--qutrit state \cite{Bound-activation}:
\begin{equation}
\rho_{\alpha}=\frac{2}{7}\left|\Psi_{+}\right\rangle \left\langle \Psi_{+}\right|+\frac{\alpha}{21}\sigma_{+}+\frac{5-\alpha}{21}\sigma_{-},
\end{equation}
where 
\begin{equation}
\left|\Psi_{+}\right\rangle =\left(\left|00\right\rangle +\left|11\right\rangle +\left|22\right\rangle \right)/\sqrt{3},
\end{equation}
is now a two--qutrit maximally entangled state and 
\begin{equation}
\sigma_{+}=\textrm{diag}\left(0,1,0,0,0,1,1,0,0\right),
\end{equation}
\begin{equation}
\sigma_{-}=\textrm{diag}\left(0,0,1,1,0,0,0,1,0\right).
\end{equation}
For $2\leq\alpha\leq3$ the state $\rho_{\alpha}$ is separable, for
$3<\alpha\leq4$, it is entangled but PPT (bound entangled), while
for $4<\alpha\leq5$ the state is entangled and not PPT.

A straightforward calculation yields 
\begin{equation}
\textrm{Tr}\rho_{\alpha}^{2}=\frac{37+2\alpha\left(\alpha-5\right)}{147}.
\end{equation}
The product numerical radius is given by the formula \cite{Gaw+10}
\begin{equation}
L\left(\rho\right)=\max_{\left|\psi\right\rangle ,\left|\chi\right\rangle }\left\langle \chi\right|\otimes\left\langle \psi\right|\rho\left|\chi\right\rangle \otimes\left|\psi\right\rangle, \label{numr}
\end{equation}
where the maximum is taken over the set of normalized states $\left\langle \psi\left|\psi\right\rangle \right.=1$ and $\left\langle \chi\left|\chi\right\rangle \right.=1$.
In order to find $L\left(\rho_{\alpha}\right)$ we parametrize $\left|\psi\right\rangle =a\left|0\right\rangle +b\left|1\right\rangle +c\left|2\right\rangle $,
with the constraint $\left|a\right|^{2}+\left|b\right|^{2}+\left|c\right|^{2}=1$ and find:
\begin{eqnarray}
L\left(\rho_{\alpha}\right) & = & \max_{\left|\chi\right\rangle }\left\langle \chi\right|\left[\frac{2}{21}\left|\psi\right\rangle \left\langle \psi\right|+\frac{\alpha}{21}\textrm{diag}\left(\left|b\right|^{2},\left|c\right|^{2},\left|a\right|^{2}\right)\right.\nonumber \\
 & + & \left.\frac{5-\alpha}{21}\textrm{diag}\left(\left|c\right|^{2},\left|a\right|^{2},\left|b\right|^{2}\right)\right]\left|\chi\right\rangle .
\end{eqnarray}
The maximization with respect to $\left|\chi\right\rangle$ gives the largest eigenvalue of the $3\times3$ matrix suited between $\left\langle \chi\right|\ldots\left|\chi\right\rangle$. This eigenvalue depends only on the moduli:  $\left|a\right|$,  $\left|b\right|$  and  $\left|c\right|$, thus due to the normalization condition it is a complicated, two--variable function. This function attains its maximum  for:
\begin{equation}
\begin{cases}
\left|a\right|=\left|b\right|=\left|c\right|=1/\sqrt{3} & \quad\textrm{for }\alpha\leq11/3\\
\left|a\right|=1\;\vee\;\left|b\right|=1\;\vee\;\left|c\right|=1 & \quad\textrm{for }\alpha\geq11/3
\end{cases}.
\end{equation}
\begin{figure}
\includegraphics[scale=0.5]{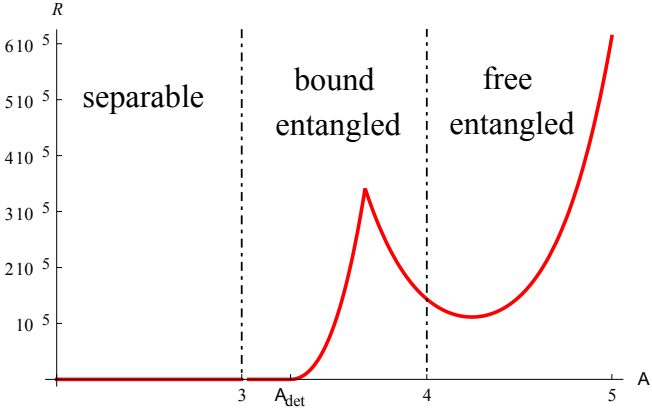}
\caption{The lower bound for $E\left(\rho_{\alpha}\right)$ as a function of $\alpha$. Positive values implies entanglement
 (bound entanglement for $\alpha\leq4$).
\label{Fig:BE}}
\end{figure}

Plugging the values of optimizing parameters we obtain the product numerical radius: 
\begin{equation}
L\left(\rho_{\alpha}\right)=\frac{1}{21}\max\left[\frac{11}{3};\alpha\right].
\end{equation}

According to our method we find the separability threshold to be
\begin{equation}
\alpha_{\textrm{det}}=\frac{15+\sqrt{21}}{6}\approx3.264.
\end{equation}
This implies that for $\alpha\in\left]\alpha_{\textrm{det}},4\right]$
the bound entanglement of $\rho_{\alpha}$ is detected. In Fig. \ref{Fig:BE}
we show the lower bound for $E\left(\rho_{\alpha}\right)$.

\section{Derivation of the upper bound for 
product numerical radius $L\left(\rho\right)$} 
Let us first recall the expansion of an arbitrary bipartite state $\rho$ given in the manuscript:
\begin{eqnarray}
\rho & = & \frac{1}{MN}\left[I_{M}\otimes I_{N}+ k_M \sum_{i=1}^{M^{2}-1}q_{i}\sigma_{i}\otimes I_{N}\right.\\
 & + & \left. k_N \sum_{j=1}^{N^{2}-1}p_{j}I_{M}\otimes\tilde{\sigma}_{j}+k_M k_N \sum_{i=1}^{M^{2}-1}\sum_{j=1}^{N^{2}-1}B_{ij}\sigma_{i}\otimes\tilde{\sigma}_{j}\right].\nonumber 
\end{eqnarray}
We use the usual normalization of the Lie groups' generators such that: 
\begin{equation}
\textrm{Tr}\sigma_{i}\sigma_{j}=2\delta_{ij},\qquad\textrm{Tr}\tilde{\sigma}_{i}\tilde{\sigma}_{j}=2\delta_{ij},
\end{equation}
and set  $k_M=\sqrt{M M_-/2}$ and  $k_N=\sqrt{N N_-/2}$ where $M_-=M-1$ and $N_-=N-1$.

In order to bound the product numerical radius (\ref{numr}) we need to 
express the two pure states to be optimized in their Bloch representation:
\begin{equation}
\rho_\chi=\left|\chi\right\rangle \left\langle \chi\right|=\frac{1}{M}\left(I_{M}+k_M\sum_{i=1}^{M^{2}-1}v_{i}\sigma_{i}\right), \label{chi}
\end{equation}
\begin{equation}
\rho_\psi =\left|\psi\right\rangle \left\langle \psi\right|=\frac{1}{N}\left(I_{N}+k_N\sum_{j=1}^{N^{2}-1}w_{j}\tilde{\sigma}_{j}\right). \label{psi}
\end{equation}
parametrized by two Bloch vectors  $\boldsymbol{v}\in\mathcal{B}\left(M\right)$,  $\boldsymbol{w}\in\mathcal{B}\left(N\right)$. Since we are looking for the upper bound on $L\left(\rho\right)$ we are allowed to relax some of the constraints defining $\mathcal{B}\left(M\right)$ and $\mathcal{B}\left(N\right)$. In our optimization routine we shall thus restrict ourselves the couple of norm constraints given by:
\begin{equation}
\sum_{i=1}^{M^{2}-1} v_{i}^{2}=1,\qquad \sum_{j=1}^{N^{2}-1} w_{j}^{2}=1.
\end{equation}
Note that in the case of qubits this couple completely characterizes the Bloch vectors. Eq. (\ref{numr}) in terms of (\ref{chi}) and (\ref{psi}) reads
\begin{eqnarray}
L\left(\rho\right) & = & \frac{1}{MN}\max_{\boldsymbol{v},\boldsymbol{w}}\left[1+N_{-}\boldsymbol{p}\cdot\boldsymbol{w}\right.\\
 & + & \left.M_{-}\boldsymbol{q}\cdot\boldsymbol{v}+M_{-}N_{-}\boldsymbol{v}\cdot\mathbf{B}\boldsymbol{w}\right].\nonumber 
\end{eqnarray}
Note that $\boldsymbol{v}\cdot\mathbf{B}\boldsymbol{w}=\boldsymbol{w}\cdot\mathbf{B}^T\boldsymbol{v}$.
We shall next maximize every part independently
\begin{eqnarray}
L\left(\rho\right) & \leq & \frac{1}{MN}\left[1+N_{-}\max_{\boldsymbol{w}}\left(\boldsymbol{p}\cdot\boldsymbol{w}\right)\right.\label{Lb}\\ 
 & + & \left.\max_{\boldsymbol{v}}\left(\boldsymbol{q}\cdot\boldsymbol{v}\right)+M_{-}N_{-}\max_{\boldsymbol{v},\boldsymbol{w}}\left(\boldsymbol{v}\cdot\mathbf{B}\boldsymbol{w}\right)\right].\nonumber 
\end{eqnarray}
We get:
\begin{equation}
\max_{\boldsymbol{w}}\left(\boldsymbol{p}\cdot\boldsymbol{w}\right)=\left\Vert \boldsymbol{p}\right\Vert,\quad\max_{\boldsymbol{v}}\left(\boldsymbol{q}\cdot\boldsymbol{v}\right)=\left\Vert \boldsymbol{q}\right\Vert,
\end{equation} where $\left\Vert \boldsymbol{x}\right\Vert=\sqrt{\sum_i x_i^2}$ and
\begin{equation}
\max_{\boldsymbol{v}}\left(\boldsymbol{v}\cdot\mathbf{B}\boldsymbol{w}\right)=\left\Vert \mathbf{B}\boldsymbol{w}\right\Vert \;\; \mathrm{ or }\;\; \max_{\boldsymbol{w}}\left(\boldsymbol{w}\cdot\mathbf{B}^T\boldsymbol{v}\right)=\left\Vert \mathbf{B}^T\boldsymbol{v}\right\Vert.
\end{equation}
Finally we maximize
\begin{equation}
\max_{\boldsymbol{w}}\left\Vert \mathbf{B}\boldsymbol{w}\right\Vert =\sqrt{\max_{\boldsymbol{w}}\left(\boldsymbol{w}\mathbf{B}^T \cdot\mathbf{B}\boldsymbol{w}\right)}=\sqrt{\xi_{1}\left(\mathbf{B}^T\mathbf{B}\right)}, \label{Eig1}
\end{equation}
or
\begin{equation}
\max_{\boldsymbol{v}}\left\Vert \mathbf{B}^T\boldsymbol{v}\right\Vert =\sqrt{\max_{\boldsymbol{v}}\left(\boldsymbol{v}\mathbf{B}\cdot\mathbf{B}^T\boldsymbol{v}\right)}=\sqrt{\xi_{1}\left(\mathbf{B}\mathbf{B}^T\right)}, \label{Eig2}
\end{equation}
where $\xi_{1}\left(\cdot\right)$ denotes the largest eigenvalue of a matrix. Both results (\ref{Eig1}) and (\ref{Eig2}) are equivalent as the spectra of $\mathbf{B}^T\mathbf{B}$ and $\mathbf{B}\mathbf{B}^T$ shall only differ by the degeneracy of the trivial eigenvalue $\xi=0$ which can equal $\xi_1$ provided that $\mathbf{B}=0$.

Plugging the above results into (\ref{Lb}) finishes the derivation of the upper bound for $L\left(\rho\right)$ given in the manuscript.

\section{Two--qubit system: Experimental implementation with $10$ parameters} 
Let us first derive the form of the $g\left(p\right)$ function for the two--qubit Werner state. We shall start recalling that we have: $C_{11}=\left(1-p\right)^{2}=C_{22}$,
$C_{12}=0$, and we perform an optimization with respect to $C_{13}$, $C_{23}$ and $C_{33} $. We also use the positivity condition
\begin{equation}
C_{13}^{2}+C_{23}^{2}\leq C_{33}\left(1-p\right)^{2}. \label{positiv}
\end{equation}
The largest eigenvalue of $\mathbf{C}$ reads
\begin{equation}
\xi_{1}\left(\mathbf{C}\right)=\frac{S_+ +\sqrt{S_-^2+4\left(C_{13}^{2}+C_{23}^{2}\right)}}{2}, \label{l1}
\end{equation}
where $S_\pm=\left(1-p\right)^{2}\pm C_{33}$.
The trace of $\mathbf{C}$  in equal to 
\begin{equation}
\textrm{Tr}\mathbf{C}=2\left(1-p\right)^{2}+C_{33}.
\end{equation}
As $\textrm{Tr}\mathbf{C}$ does not depend on $C_{13}$ and $C_{23}$ we perform the optimization with respect to these variables applying the positivity condition (\ref{positiv}) to (\ref{l1})
\begin{equation}
\max_{C_{13},C_{23}}\xi_{1}\left(\mathbf{C}\right)=\left(1-p\right)^{2}+C_{33}\equiv S_+.
\end{equation}
Finally we get the desired result:
\begin{eqnarray}
g\left(p\right)& = & \max_{C_{33}}\left(\sqrt{\left(1-p\right)^{2}+C_{33}}-2\left(1-p\right)^{2}-C_{33}\right)\nonumber\\
 & = & \begin{cases}
2\left(1-p\right)\left(p-\frac{1}{2}\right) & \qquad\textrm{for }0\leq p\leq\frac{1}{2}\\
\frac{1}{4}-\left(1-p\right)^{2} & \qquad\textrm{for }\frac{1}{2}\leq p\leq1
\end{cases}.
\end{eqnarray}

\subsection{Relations between parameters} 
Denote by $\left|\phi_{\pm}^{k}\right\rangle $, $k\in\left\{ 1,2,3\right\} $
two eigenstates of the Pauli matrix $\sigma_{k}$ such that $\sigma_{k}=\left|\phi_{+}^{k}\right\rangle \left\langle \phi_{+}^{k}\right|-\left|\phi_{-}^{k}\right\rangle \left\langle \phi_{-}^{k}\right|$
and introduce the following six $2\times2$ matrices:\begin{eqnarray}
\Omega_{\pm}^{k} & = & \textrm{Tr}_{B}\left(\rho I_{A}\otimes\left|\phi_{\pm}^{k}\right\rangle \left\langle \phi_{\pm}^{k}\right|\right)\\
 & \equiv & \frac{1}{4}\left[\left(1\pm p_{k}\right)I_{A}+\sum_{j=1}^{3}\left(q_{j}\pm B_{jk}\right)\sigma_{j}\right].\nonumber 
\end{eqnarray}
With the help of the above matrices we define $15$ parameters which
via quantum tomography completely describe the state $\rho$. We
group these parameters in four families:
\begin{enumerate}
\item three traces ($k\in\left\{ 1,2,3\right\} $)
\begin{equation}
\mathcal{T}_{k}=\textrm{Tr}\Omega_{+}^{k}\equiv\frac{1+p_{k}}{2},\label{tr}
\end{equation}

\item six norms ($k\in\left\{ 1,2,3\right\} $)
\begin{equation}
\mathcal{P}_{k}^{\pm}=\textrm{Tr}\left(\Omega_{\pm}^{k}\right)^{2}\equiv\frac{1}{8}\left[\left(1\pm p_{k}\right)^{2}+\sum_{j=1}^{3}\left(q_{j}\pm B_{jk}\right)^{2}\right],
\end{equation}

\item three overlaps ($\left(k,l\right)\in\left\{ \left(1,2\right),\left(1,3\right),\left(2,3\right)\right\} $)
\begin{eqnarray*}
\mathcal{F}_{kl} & = & \textrm{Tr}\left(\Omega_{+}^{k}\Omega_{+}^{l}\right)\\
 & \equiv & \frac{1}{8}\left[\left(1+p_{k}\right)\left(1+p_{l}\right)+\sum_{j=1}^{3}\left(q_{j}+B_{jk}\right)\left(q_{j}+B_{jl}\right)\right],
\end{eqnarray*}

\item three cross terms
\begin{equation}
\mathcal{G}_{k}=\textrm{Tr}\left(\Omega_{+}^{k}\Omega_{-}^{k}\right)\equiv\frac{1}{8}\left[1-p_{k}^{2}+\sum_{j=1}^{3}\left(q_{j}^{2}-B_{jk}^{2}\right)\right].\label{cross}
\end{equation}
\end{enumerate}

We shall now derive the relation between the quantities $\left\Vert \boldsymbol{p}\right\Vert^2 $,
$\left\Vert \boldsymbol{q}\right\Vert^2 $, $C_{11}$, $C_{22}$, $C_{12}$
and the parameters (\ref{tr}-\ref{cross}). We find that $p_{k}=2\mathcal{T}_{k}-1$
so 
\begin{equation}
\left\Vert \boldsymbol{p}\right\Vert ^{2}=\sum_{k=1}^{3}\left(2\mathcal{T}_{k}-1\right)^{2},\label{p}
\end{equation}
and:
\begin{eqnarray}
\left\Vert \boldsymbol{q}\right\Vert ^{2} & = & 4\mathcal{G}_{1}+2\left(\mathcal{P}_{1}^{+}+\mathcal{P}_{1}^{-}\right)-1\\
 & \equiv & 4\mathcal{G}_{2}+2\left(\mathcal{P}_{2}^{+}+\mathcal{P}_{2}^{-}\right)-1,\nonumber 
\end{eqnarray}
\begin{equation}
C_{11}=4\left(\mathcal{P}_{1}^{+}+\mathcal{P}_{1}^{-}\right)-\left(2\mathcal{T}_{1}-1\right)^{2}-\left\Vert \boldsymbol{q}\right\Vert ^{2}-1,
\end{equation}
\begin{equation}
C_{22}=4\left(\mathcal{P}_{2}^{+}+\mathcal{P}_{2}^{-}\right)-\left(2\mathcal{T}_{2}-1\right)^{2}-\left\Vert \boldsymbol{q}\right\Vert ^{2}-1,
\end{equation}
\begin{eqnarray}
C_{12} & = & 8\mathcal{F}_{12}-\left(2\mathcal{T}_{1}-1\right)\left(2\mathcal{T}_{2}-1\right)-\left\Vert \boldsymbol{q}\right\Vert ^{2}\nonumber \\
 & - & 1-2\left(\mathcal{P}_{1}^{+}-\mathcal{P}_{1}^{-}\right)-2\left(\mathcal{P}_{2}^{+}-\mathcal{P}_{2}^{-}\right).
\end{eqnarray}

As mentioned in the main paper to determine the degree of entanglement
on an arbitrary two--qubit mixed state it is sufficient to measure only $10$ parameters: 
\begin{equation}
\mathcal{T}_{1},\mathcal{T}_{2},\mathcal{T}_{3},\quad\mathcal{G}_{1},\mathcal{G}_{2},\quad\mathcal{P}_{1}^{+},\mathcal{P}_{1}^{-},\mathcal{P}_{2}^{+},\mathcal{P}_{2}^{-},\quad\mathcal{F}_{12},
\end{equation}
which is less than $15$ required by the standard quantum tomographic procedure.
\end{document}